# A theoretical interpretation of variance-based convergence criteria in perturbation-based theories

Xiaohui Wang[1] and Zhaoxi Sun[1,2*]

[1]*State Key Laboratory of Precision Spectroscopy, School of Chemistry and Molecular Engineering, East China Normal University, Shanghai 200062, China*

[2]*Computational Biomedicine (IAS-5/INM-9), Forschungszentrum Jülich, Jülich 52425, Germany*

*To whom correspondence should be addressed: z.sun@fz-juelich.de

## Abstract

Free energy simulation at high level of theory can be performed indirectly by constructing the thermodynamic cycle, where a low level force-field free energy simulation is combined with high level end-state corrections. As huge differences exist between the computational cost of low level MM and high level QM Hamiltonians, the errors in QM/MM indirect free energy simulation are mostly introduced in QM/MM end-state corrections. As a specific type of alchemical free energy simulation, QM/MM corrections can be obtained from integration of the partial derivatives of alchemical Hamiltonians or from perturbation-based estimators including free energy perturbation (FEP) and acceptance ratio methods. When using FEP or exponential averaging (EXP), a number of researchers tend to only simulate MM end states and calculate single point energy in order to get the free energy estimates. In this case the sample size hysteresis arises and the convergence is determined by bias elimination rather than variance minimization. Various criteria are proposed to evaluate the convergence issue of QM/MM corrections and numerical studies are reported. It has been observed that criteria including the variance of distribution, the effective sample size, information entropies and so on can be used and they are somehow variance-of-distribution-dependent. However, no theoretical interpretation has been presented. In this paper we present theoretical interpretations to dig the underlying statistical nature behind the problem. Those convergence criteria are proven to be related with the variance of the distribution in Gaussian approximated Exponential averaging (GEXP). Notably, for the first time, we prove that these estimators are nonlinearly dependent on the variance of the free energy estimate. As these estimators are often orders of magnitude overestimated, the variance of the FEP estimate is orders of magnitude underestimated. Hence, computing this statistical uncertainty is meaningless in practice unless convergence is already achieved. In numerical calculation from timeseries data the effective sample size is bounded by 1 and N and thus the variance of the free energy estimate is proven to be bounded by 0 and $1(k_BT)^2$ for EXP and 0 and $2(k_BT)^2$ for BAR (or the standard deviation of EXP is smaller than $k_BT$ and that of BAR is smaller than $\sqrt{2}k_BT$), which indicates an inevitable underestimation. To be specified, the upper bounds for these estimators are sample-size dependent. The effective sample size is also proven to be a function of the overlap scalar, from which the range of the overlap scalar can also be derived. These findings deepen our understanding of perturbation-based theories.

# I. Introduction

Free energy difference between different states or models of a given system determines the thermodynamic tendency of physical processes in various scientific fields.[1-10] To obtain a reliable free energy estimate from free energy simulation, adequate sampling and accurate Hamiltonians are required. The sampling issue can solved via enhanced sampling methods such as slow growth,[11-13] linear response approximation,[14-15] linear interaction energy[15-18] and expanded ensemble simulations.[19-25] Nonequilibrium scenarios are introduced by Jarzynski[26-27] and Crooks,[28] which provide alternative simulation protocols to enhance the sampling efficiency. With enhanced sampling methods and proper reweighting, we extend the molecular dynamics (MD) accessible timescale significantly and converged statistics can be obtained. Then we can check the reliability and accuracy of Hamiltonian we used.

Although the errors associated with force fields are introduced in parameterization and are unable to quantify exactly, molecular mechanics (MM) Hamiltonians or even coarser models enable milliseconds (to microseconds) simulation, thus making direct observations of phenomena happening at experimental timescales possible. A more detailed description is always desirable due to improved accuracy. However, quantum mechanics (QM) and various multiscale QM/MM treatments are computationally demanding for well-converged phase space sampling and even prohibitive for biomolecules.[29-31] Indirect QM/MM free energy simulations are thus introduced, which find alternative transformation pathways by constructing a thermodynamic cycle connecting two QM/MM end states.[32-36] Relatively cheap free energy simulations at MM level and sometimes at semi-empirical QM level (SQM) are combined with end-state QM/MM corrections to reproduce the thermodynamics quantities in the direct QM/MM transformation. QM end-state corrections are often implemented with Thermodynamic Integration (TI) or perturbation-based theories.[37-47] Free energy profiles can also be obtained in a similar 'indirect' way.[48-52] If the potential energy surfaces of MM and QM descriptions are significantly different, the overlap of the phase spaces under QM and MM Hamiltonians is small and configurations sampled at MM level would not be representative of those at QM level.[53] A broadly distributed energy differences are expected and a strong bias exists in the free energy estimate, triggering a poor convergence behavior. To minimize the gap with a balanced computational cost, efforts are mainly on a) inserting a medium level Hamiltonian between MM and QM as intermediate states followed by the normal windows sampling workflow, or simply replacing MM with SQM calculations,[53] or modifying the MM Hamiltonian to get a better description as well as a better phase space overlap, b) using different transformation scheme such as non-equilibrium methods of Jarzynski's Identity (JI) and Crooks' Equation (CE).[27-28, 54] Although there are a number of bidirectional estimators that are statistically efficient and asymptotically unbiased, due to the orders of magnitude differences between computational costs of QM and MM Hamiltonians, single state simulations using FEP-type estimators (including FEP and JI) are always preferable, even it is often not so reliable. In this case sample size hysteresis arises.[55-57] The numerical stability of exponential average is bad and abrupt changes happen. Large bias often exists in the free energy estimate. Thus researches on the sample size required to achieve convergence often concentrate on bias elimination, namely the number of sample needed to get an unbiased FEP result, rather than minimizing the variance or uncertainty of the free energy estimate.

Very recently, Ryde presented a numerical study on the sample size issue and compared several criteria for the convergence of EXP estimates under Gaussian approximation.[58] This study, focusing on Gaussian distributed energy difference in QM/MM corrections, did provide critical insights into how many configurations are needed for a converged EXP estimate, GEXP estimate and the performances of various convergence criteria to determine the convergence of EXP. The exponential dependence of the number of samples needed to obtain a converged EXP estimate on the standard deviation of the distribution of energy difference was observed. The information entropy was observed to be negatively correlated with the variance of the distribution. He also concluded the variance of the distribution of energy difference was the best convergence criterion. However, a theoretical explanation of these phenomena was absent. The paper only presented a numerical solution and practical considerations of the convergence issue. A theoretical interpretation was thus called for.

In this paper, firstly, also under Gaussian approximation, we take one step further by presenting theoretical derivations and proving the consistency between two convergence criteria including the effective sample size and the information entropy, as is shown in Eq. (9-12). A theoretical interpretation of the exponential dependence of the sample size on the variance of the distribution and the negative correlation between the information entropy and the variance of the distribution are analytically derived as Eq. (9-12, 19, 20). We offer an explanation of why the variance of the distribution is the central quantity among all convergence criteria in Gaussian approximated exponential averaging: the variance of the distribution appears in the analytical formulas of these convergence criteria. Also, the numerical calculation of the effective sample size with Eq. (16) offers an overestimation over the analytical formula of Eq. (9).

Secondly, to get further insights of the convergence criteria, relating them with the ones universally used such as the variance of the free energy estimate is always preferable. The effective sample size has been observed to be negatively correlated with the variance of the free energy estimate in a numerical study.[59] Hence, we then derive the relationship between the effective sample size and the variance of the free energy estimate in perturbation based theories (including FEP and BAR) as Eq. (21, 22, 27, 28), and thus provide a different point of view of the variance of the free energy estimate. The Gaussian approximated equation is derived. Here, we would also like to jump out of the Gaussian approximation and provide more general relationships. Specifically, Eq. (21) finds the general relationship between the variance of the free energy difference and the effective sample size and Eq. (22) offers a new way to derive the formula of GEXP variance in Eq. (5b). From the range of the effective sample size, we find a new and extremely significant property of the variance of the free energy estimate: the variance of EXP estimate is always smaller than 1 $(k_B T)^2$ and that of BAR estimate is always smaller than 2 $(k_B T)^2$, as is shown in Eq. (23a, 29). To be specified, the upper bounds are smaller than 1 $(k_B T)^2$ or 2 $(k_B T)^2$ and are sample-size dependent, as is summarized in Eq. (33). Thus the standard deviation of EXP is smaller than $k_B T$ and that of BAR is smaller than $\sqrt{2} k_B T$. Under model approximation (Gaussian distributed energy difference), there is no upper bound for GEXP, as is given in Eq. (23b). Further, according to the overestimation of the effective sample size, the variance of the free energy estimate is significantly underestimated.

Finally, as a number of convergence criteria such as time-derivatives of the variance (TDV)[60-62] and the overlap estimators[63-64] are dependent on the variance of the free energy estimate, we prove that they are also non-linear functions of the effective sample size, as is shown in Eq. (30, 31). More importantly, the range of the overlap scalar obtained from timeseries data is derived from its relationship with the variance or the effective sample size or from its definition below Eq. (30), as is shown in Eq. (32). The sample-size dependent range of the overlap scalar indicates that to obtain reliable estimate of the phase space overlap, especially for small overlap cases, a large number of samples are required.

## II. Convergence criteria under Gaussian Approximation in Exponential Averaging

**Gaussian distributed energy difference and linear response in QM/MM correction.** QM/MM corrections are essentially electrostatic corrections, as the vdW parts of QM and MM Hamiltonians are normally the same. (We do not consider dispersion corrections here.) Thus, using linear mixing of Hamiltonians, according to Linear Response Approximation (LRA),[14] the electrostatic response should be linear, namely $\left\langle \frac{\partial U}{\partial \lambda} \right\rangle_\lambda$ varying linearly with the order parameter λ. Here $U(\lambda) = \lambda U_j + (1-\lambda)U_i$ is the linearly mixed reduced energy function and $U_i$ is the reduced energy function in state i. Considering this feature, to perform QM/MM correction, TI with LRA is a reasonable choice. However, as TI with LRA requires computational demanding simulation in QM/MM state, to reduce the computational cost, researchers tend to use methods that only requires simulations at MM state. In LRA the distribution of the energy difference is Gaussian. Hence, normally the QM/MM correction is performed via FEP with Gaussian approximations.

**Consistency of convergence criteria under Gaussian Approximation.** Firstly we define

$$x = \Delta U(\mathbf{q})_{ij} = U(\mathbf{q})_j - U(\mathbf{q})_i \tag{1}$$

, as the dimensionless energy difference between state i and state j for sample with coordinate vector $\mathbf{q}$. In the formulism of EXP, the dimensionless free energy difference between state i and state j $\Delta F_{ij}$ and the corresponding variance can be calculated by

$$\Delta F_{ij} = F_j - F_i = -\ln\left(\frac{1}{N_i}\sum_{n=1}^{N_i} e^{-\Delta U(\mathbf{q}_n)_{ij}}\right) = -\ln\left\langle e^{-x} \right\rangle_i \tag{2a}$$

$$\sigma^2_{\Delta F_{ij}} = \frac{\sigma^2_{e^{-x},i}}{N_i \left\langle e^{-x} \right\rangle_i^2} \tag{2b}$$

, where n refers to the nth sample from state i, $N_i$ is the sample size in state i, $\sigma^2_{e^{-x},i}$ denotes the variance of $e^{-x}$ computed with samples from state i, and conventional angle brackets represent canonical averages.[65-67] Under Gaussian approximation, the distribution of $x$ follows

$$\rho(x) = \frac{1}{\sqrt{2\pi}\sigma} e^{-\frac{(x-\mu)^2}{2\sigma^2}} \tag{3}$$

, where $\sigma$ is the standard deviation of the energy difference and the center of the distribution is at $\mu$.

We hereafter omit subscripts i and j without causing confusion. Under this approximation, we have

$$\langle e^{-x} \rangle = \int_{-\infty}^{\infty} e^{-x} \rho(x) dx = e^{\frac{\sigma^2}{2} - \mu} \tag{4}$$

. The GEXP estimate can thus be obtained with

$$\Delta F = \mu - \frac{\sigma^2}{2} \tag{5a}$$

$$\sigma_{\Delta F}^2 = \frac{\sigma^2}{N} + \frac{\sigma^4}{2(N-1)} \tag{5b}$$

.[66-67] The weight of sample n in the exponential average is

$$w_n = \frac{e^{-x(\mathbf{q}_n)}}{\sum_{n=1}^{N} e^{-x(\mathbf{q}_n)}} \tag{6}$$

. Following Ryde we consider convergence criteria of Kish's approximate estimation of the effective sample size

$$Q = \frac{\left(\sum_{n=1}^{N} w_n\right)^2}{\left(\sum_{n=1}^{N} w_n^2\right)} \tag{7}$$

, and the information entropy

$$S = -\frac{1}{N} \sum_{n=1}^{N} (w_n \ln w_n) \tag{8}$$

.[58, 68] Applying the Gaussian approximation into numerical integration, in the large sample size regime, we have

$$Q_{GEXP} = \frac{(N\langle w \rangle)^2}{N\langle w^2 \rangle} = \frac{N}{e^{\sigma^2}} \tag{9}$$

, and the entropy becomes

$$S = -\langle w \ln w \rangle = \frac{-\frac{\sigma^2}{2} + \ln N}{N} \tag{10}$$

. Now, an obvious phenomenon can be seen. These convergence criteria are actually calculating the same thing—$\sigma^2$, the variance of the distribution of the energy difference. This is an important value which is linearly dependent on the GEXP estimate in Eq.(5a) and needs to be converged in GEXP. Hence, these convergence criteria are actually evaluating the convergence of the distribution of the energy difference and their results are essentially the same. We can rescale S by its information length for normalization

$$S_1 = \frac{NS}{\ln N} = \frac{-\frac{\sigma^2}{2} + \ln N}{\ln N} \tag{11}$$

, or multiply it by N to make it more similar with the logarithm of $Q$ and recover the Shannon entropy of

$$S_0 = NS = -\frac{\sigma^2}{2} + \ln N \tag{12a}$$

$$\ln Q_{GEXP} = \ln N - \sigma^2 = S_0 - \frac{\sigma^2}{2} \tag{12b}$$

. Obvious conclusions from this equation are: a) $\ln Q_{GEXP}$ is more sensitive to small variations in the variance of the distribution of the energy difference than the entropy, namely

$$\frac{\partial \ln Q_{GEXP}}{\partial \sigma} = 2 \frac{\partial S_0}{\partial \sigma} = -2\sigma \tag{13}$$

, b) their responses to the number of samples are the same,

$$\frac{\partial \ln Q_{GEXP}}{\partial N} = \frac{\partial S_0}{\partial N} = \frac{1}{N} \tag{14}$$

. These two conclusions are only valid when the exponential average is converged. If not, we expect the curves to vary from the ideal case. Hence, from a theoretical point of view, to get statistics that define the convergence, we should always use Kish's effective sample size or its logarithm rather than the entropy.

**The converged curvature of the $Q-N$ curve indicates converged GEXP estimates.** When $Q=1$, $N=e^{\sigma^2}$, meaning that the number of independent samples required to get an effective sample increases exponentially with the variance of the distribution. As in using these convergence criteria we always check their time-evolution behaviors and find meaningful variations to determine convergence,[58] we check the derivatives of the effective sample size

$$\frac{\partial Q_{GEXP}}{\partial N} = e^{-\sigma^2} \tag{15a}$$

$$\frac{\partial Q_{GEXP}}{\partial \sigma} = -2N\sigma e^{-\sigma^2} \tag{15b}$$

. Eq. (15a) shows that when the linear dependence of $Q_{GEXP}$ on the sample size is achieved, converged $\sigma$ is obtained and thus the converged free energy estimate is obtained, according to the formula of GEXP in Eq.(5). From Eq. (15b) we notice that the sensitivity of $Q_{GEXP}$ to small variations of $\sigma$ varies with N and $\sigma$. Hence, the magnitude of $\sigma$ should also be taken into consideration. Monitoring the slope of the $Q-N$ curve and considering the magnitude of $\sigma$ (the sensitivity of $Q$ to the change of $\sigma$) we can somehow determine whether the convergence is achieved. Still, we should note that the above relation is only valid in the large sample size regime.

A generalization of GEXP $Q$ in Eq.(9) is the EXP formula,

$$Q_{EXP} = \frac{N\langle e^{-x}\rangle^2}{\langle e^{-2x}\rangle} \tag{16}$$

, which means that $Q$ is an estimate of exponential average. Hence, the curvature of $Q-N$ provides direct estimates of those exponential terms. Converged slopes indicate converged exponential averages. From the sample size hysteresis we know that if $\sigma$ is too large the abrupt changes can be seen in the timeseries of these exponential terms due to the observations of rare events. Then, how should we define convergence with $Q$ practically? A reasonable solution is comparing the analytical line of Eq. (9) with the numerical one of Eq. (16). However, in the following parts we show that those variance based convergence criteria are generally useless in the exponential averaging case.

Before proceeding, about the range of the effective sample size we should give some discussions. According to the definition of the effective sample size $Q$ in Eq. (7, 16), its value is within the range between 1 and N. However, the analytical formula for Gaussian model in Eq. (9) indicates an upper bound of N but a lower bound of 0. These are summarized as

$$Q_{EXP} \in [1, N] \tag{17a}$$

$$Q_{GEXP} \in [0, N] \quad (17b)$$

. If the Gaussian distribution is exactly followed, although both Eq. (9) and Eq. (16) are asymptotically unbiased, the analytical result is more statistically efficient than the numerical one. This means that the effective sample size can be smaller than 1. In the small sample size case where the analytical result is smaller than 1, the numerical result obtained from Eq. (7, 16) is inevitably overestimated, and no converged result can be obtained. Later we also relate the effective sample size with the variance of the free energy estimate and we can see that overestimated $Q$ leads to underestimated variance and the bounds of the numerical $Q$ lead to bounds of the variance of the free energy estimate obtained from timeseries data.

### III. Numerical problems of the convergence criteria in Exponential Averaging

**Simulation details.** As we are dealing with ideal Gaussian distributions, timeseries data are generated from model MD simulation with harmonic potential energy function to get the desired Gaussian distribution. Statistical inefficiency is computed to extract independent samples.[69-73] In the following parts we only deal with uncorrelated samples. A comparison between distributions from numerical samples and the analytical formula is shown in Figure 1.

As at 300 K for dimensionless $\sigma$ 1 equals 0.59 kcal/mol (2.46 kJ/mol) and 10 equals 5.9 kcal/mol (24.6 kJ/mol), we consider $\sigma$ as integers from 1 to 10, which already covers the normal range of $\sigma$ observed in the free energy perturbation. An illustration of these Gaussian distributions is shown in Figure 1.

**Is Q really useful as a convergence criterion in exponential averaging?** Firstly in Figure 2 we present the theoretical $Q_{GEXP}$ surface in $\sigma$ and N spaces, its projections on $\sigma$ and its derivatives. As has been pointed out above, according to the definition of the effective sample size, its value is within the range between 1 and N. Hence the curves are truncated at $Q = 0.1$. From this plot we notice that theoretically the effective sample size can be smaller than 1, which can never be achieved with the numerical calculation via Eq. (16). This indicates the existence of an overestimation in these regions. From $\sigma = 0$ $Q$ decreases exponentially with $\sigma$, and for $\sigma$ as large as 4 we already cannot obtain one effective sample even with more than one million independent samples.

Practically, the normal sample size obtained in MD simulation ranges from 1000 to 10000, and the threshold to obtain one effective sample for $\sigma$ is about 3. If the statistical inefficiency for the central collective variable x is 10 ps, this means more than 10 ns simulation and also a large number of QM/MM single point calculations, and the outcome is only 1 sample. We thus wonder, does this worth it? From the $\frac{\partial Q_{GEXP}}{\partial \sigma}$ we notice that for $\sigma$ larger than 4, the sensitivity of $Q_{GEXP}$ almost does not change with small variations of $\sigma$. The N-dependence of $Q_{GEXP}$ is linear.

For practical considerations of using the effective sample size, in the previous section we give one reasonable solution that monitoring the time-dependent behaviors of the analytical curve ($Q_{GEXP}$ in Eq.(9)) and the numerical one ($Q_{EXP}$ in Eq.(16)) may be useful. The slope of the analytical line is an estimate of $e^{-\sigma^2}$ which determines the accuracy of GEXP estimate while the numerical curve plots the change of the exponential terms, namely the time-evolution of the EXP estimate. We show in Figure 3 the comparison between those $Q-N$ curves and in Figure 4 the convergence behaviors of EXP and GEXP estimates, in one exponential average trial under each $\sigma$.

In Figure 3 the numerical $Q-N$ curves are obtained from the timeseries data with Eq. (16) and the analytical curves are obtained with Eq. (9). Theoretical estimates are obtained from the $\sigma$ used to generate the Gaussian distributions. When $\sigma = 0$ all three estimates are identical. With increased $\sigma$ mismatches between them increase dramatically. The analytical line almost overlaps with the theoretical one while the numerical one deviates significantly. The deviation is small in the case of $\sigma = 1$ and 2 and becomes large since $\sigma = 3$. When $\sigma > 4$ theoretically we cannot even see any effective sample within 100000 samples (meaning 1 $\mu s$ sampling). By contrast, the numerical result is significantly overestimated and there are a number of effective samples in the small sample size regime.

Both the theoretical lines and the numerical lines seem to be converged. Even when $\sigma$ is small the differences between the numerical slopes and the analytical (and theoretical) slopes are relatively large. Hence, we need to identify the slope of which line is reliable and why this happens. For a real number k we know that

$$\langle e^{-kx} \rangle = e^{\frac{k^2}{2}\sigma^2 - k\mu} \tag{18a}$$

$$Q_{e^{-kx}} = \frac{N\langle e^{-kx}\rangle^2}{\langle e^{-2kx}\rangle} = \frac{N}{e^{k^2\sigma^2}} \quad (18b)$$

. The larger the absolute value of k is, the harder the convergence of the $Q_{e^{-kx}} - N$ slope will be. Also, as we know, the higher the order k is, the larger the overestimation of $Q_{\langle e^{-kx}\rangle}$ is. We can expect that the case that k equals 2 converges worse than the case that k equals 1. Thus the convergence of the slope of $Q_{EXP}$ is hard due to its dependence on higher-order terms. The fake convergence of the numerical $Q-N$ slope can be quite misleading. We cannot get an unbiased estimate of $e^{-\sigma^2}$ and thus $\sigma^2$ from the slope of the curve.

As for the overestimation of the effective sample size, an illustrative example is the case $\sigma = 5$. At the end of 100000 samples the EXP estimate is -11.623, the GEXP estimate is -12.498 and the theoretical result is -12.5. The EXP bias is 0.88, $\langle e^{-x}\rangle$ is $e^{0.88} = 2.4$ fold of the converged result and $\langle e^{-2x}\rangle$ is about $1.1\times 10^8$ smaller than the converged result, leading to a $2\times 10^7$-fold overestimation of $Q_{EXP}$. Hence, the absolute value of $Q$ can neither be used to determine the convergence.

Such overestimation of $Q$ and fake convergence of the slope tell us that those $Q$-like convergence criteria perform badly in exponential averaging. Hence, we should not use $Q$ or related terms to check the convergence. To get converged $\sigma^2$ and thus GEXP estimate monitoring the convergence of the mean and the variance of the distribution should be much more efficient (considering their formulas of $Q$).

Sudden changes in the value of $Q_{EXP}$ can be seen, which is often observed in the time-evolution of EXP estimates and indicates poor convergence. The poor convergence behavior of EXP curves in Figure 4 agrees with the poor convergence of the slope of the effective sample size. By contrast, the convergence of GEXP estimates is quite well, which should be caused by the efficiencies of ordinary average and $\sigma$ estimation. Again, rather than using $Q_{EXP}$, monitoring the convergence behavior of the standard deviation of the distribution is a better choice. Note that when $\sigma$ is small using EXP is still reliable and efficient, due to the relative largeness of $Q_{EXP}$ and $Q_{\sigma^2}$ ($e^{\sigma^2}$ and 3).

Information entropies ($S$, $S_0$, and $S_1$) show the same behaviors of the effective sample size (results not shown).

**How many configurations are needed to converge? A theoretical interpretation.** Now, we consider the number of samples needed to achieve a certain accuracy and confidence, as has been studied numerically in reference.[58] We use the exactly the same threshold (for accuracy and confidence) with the reference for comparison.

As has been observed in reference,[58] from numerical results in Figure 5 we know the exponential dependence of N on $\sigma$. We further consider the effective sample size in the ideal case,

$$\ln N = \ln Q_{GEXP} + \sigma^2 \tag{19}$$

. Ideally the logarithm of the number of independent samples should vary linearly with the variance of the distribution. We assume the numbers of effective samples needed to converge for distributions with different $\sigma$ are approximately the same, the validity of which will be discussed below Eq. (23). Based on the above equation, we plot the dependence of the number of samples for convergence on $\sigma^2$ in Figure 5. Here, a well-behaved linear dependence can be seen. In this way, when the convergence threshold is fixed, we can interpolate the number of samples needed for converged EXP estimate in large $\sigma$ cases from data obtained in small $\sigma$ cases.

As in numerical simulation the convergence is defined by the required accuracy and confidence, a convergence threshold related factor determining the slope of $\ln N - \sigma^2$ curves needs to be added,

$$\ln N = \left(\ln Q_{GEXP} + \sigma^2\right) f(threshold) \tag{20}$$

, where the threshold depends mainly on the accuracy requirement as confidence level should at least be 95% for safety. Theoretically we know that in the tightest threshold limit $f(threshold) \to \infty$ and if no accuracy is required $f(threshold) \to 0$. To quantify the threshold-dependent factor, nonlinear functions can be proposed or fitted from numerical scanning the magnitude of the convergence threshold. However, using a linear fit is simpler and easier to calculate. In Figure 6 the linear fit does not provide significantly worse estimates of the slope compared with the exact result, in the neighborhood of the accuracy required in the normal cases. Normally the accuracy of dimensionless threshold required is about 1 to 2. Hence, the slope is generally 0.4-0.45. We recommend the users use higher values such as 0.45 or 0.5 for safety. The convergence of GEXP estimate is rather simple, as it only contains the convergence of the mean and the variance of the distribution and related contents can be found in a number of references and literatures.

We should note that our findings about number of trails needed for convergence are, to some extent, the near-equilibrium perturbation limit of previous results obtained by Christopher Jarzynski in the framework of nonequilibrium free energy simulation.[74-75] Specifically, they are Jarzynski's result in the case of the fluctuation-dissipation relation.[55, 76]

IV. Variance based convergence criteria in perturbation-based theories

In the previous section, we discuss the convergence criteria and their performance in Gaussian approximated energy difference cases. However, normally the data may not follow Gaussian distribution perfectly. Hence, in the following section, we would like to jump out of the Gaussian approximation and provide more general relationships between the convergence criteria and other statistical quantities.

Aside from the criteria for single-state estimator of FEP, various criteria evaluating the convergence issue of bidirectional estimators[64, 77] and multistate estimators[78-79] are also 'variance'-dependent. Here, the 'variance' can be referred to either the variance of distribution or the variance of the free energy estimates, as they are highly correlated quantities. Take the 'variance' as the variance of the free energy difference and examples of such convergence-check quantities are TDV (linearly dependent on the variance),[60-62] the overlap matrix[63] (non-linearly dependent) and the overlap of distributions[64] (non-linearly dependent). From a statistical point of view these estimators and convergence criteria perform much better than single-directional EXP, as they are using better weighting functions than the exponential average. Specifically, in QM/MM corrections, if the Gaussian approximation is applicable, the numerical problem in EXP can be solved by applying GEXP. If not, considering the exponential dependence of the number of independent samples on the variance of the energy-difference distribution to obtain converged EXP estimates, performing relatively expensive QM/MM simulation to get information in backward perturbations (MM-> QM as forward) in order to eliminate the bias effectively should be considered.

To provide further insights into the effective sample size, we relate it with the variance of the EXP estimate. From the variance of EXP in Eq. (2b) we know direct calculation of the variance of the EXP estimate can be obtained from $Q_{EXP}$, namely

$$\sigma^2_{\Delta F, EXP} = \frac{\sigma^2_{e^{-x}}}{N\langle e^{-x}\rangle^2} = \frac{\langle e^{-2x}\rangle}{N\langle e^{-x}\rangle^2} - \frac{1}{N} = \frac{1}{Q_{EXP}} - \frac{1}{N} \qquad (21)$$

. Under Gaussian approximation the above equation becomes

$$\sigma^2_{\Delta F, GEXP} = \frac{1}{Q_{GEXP}} - \frac{1}{N} = \frac{e^{\sigma^2}-1}{N} \qquad (22)$$

. Then we see that another way to derive the GEXP variance in Eq. (5b) is cumulant expanding the above equation to the 2$^{nd}$ order term.

As has been shown above in Eq. (17), the numerical estimate in Eq. (7, 16) is bounded by 1 and N while the analytical formula in Eq. (9) is bounded by 0 and N. This indicates that variance estimates from Eq. (2b, 21) are always smaller than 1, while the analytical formula of Eq. (5b, 22) does not give such a constraint. These can be summarized as

$$\sigma^2_{\Delta F, EXP} \in [0,1) \qquad (23a)$$

$$\sigma^2_{\Delta F, GEXP} \in [0,\infty) \qquad (23b)$$

. Recalling the discussion in the previous section that the effective sample size is overestimated by orders of magnitudes, we know that the variance of the free energy estimate is significantly underestimated. An illustrative calculation for showing the existence of the upper bound of EXP variance is performed and the timeseries of EXP variance is shown in Figure 7. Considering the fact that in Figure 4 at the end of 100000 samples the EXP bias is still much larger than 1 (e.g. for distribution $\sigma > 6$), the variance cannot really reflect the degree of convergence. Thus any finite-time simulations tend to underestimate the statistical uncertainty. In the appendix, we offer another way to derive the above result (from the expansion of the ensemble average).

Normally the number of independent samples N is a large number and the variance of the free energy estimate is approximately the reciprocal of the effective sample size. Then some further discussions on the questionable results in the previous researches due to the overestimation of $Q_{EXP}$ should be done. Take the result in ref. 37 as an example. In its Figure 6c, when the desired accuracy (bias elimination) is achieved, such as 4 kJ/mol ($\approx 1.63$), the standard error of the EXP estimate should be larger than the bias should be larger than the bias 1.63 in order to get the reliably unbiased estimate. Thus the effective sample size should be smaller than about 0.38. However, we can see two orders of magnitude overestimation of $Q_{EXP}$ in this result, which can never be called convergence. Further, as the slope of $Q_{EXP}$ is extremely hard to converge and so is the EXP estimate, the variance or standard error of the EXP estimate is orders of magnitude underestimated. Hence, we deeply question the meaning of calculating these error quantities. If the distribution is not Gaussian, this phenomenon is further exaggerated. We therefore ask, do we actually need those extremely underestimated values to determine convergence?

Another thing we can interpret is the number of effective samples to get a converged EXP estimate. As is assumed around Eq. (19), the converged EXP estimate requires almost the same effective sample size. According to Eq. (21), this means that the variance of the EXP estimate should be decreased to a certain threshold. This is what we normally expect to get a converged (unbiased) EXP estimate.

If we use the Fermi function to weight samples and include the optimal shift in weighting as is done in BAR[64, 77] (or more generally the optimal weighting in the extended bridge sampling regime),[80] the free energy difference and the corresponding variance can be expressed as

$$\begin{cases} \Delta F = \ln \frac{\langle f(C-x) \rangle_j}{\langle f(x-C) \rangle_i} + C \\ C = \Delta F + \ln(\frac{N_j}{N_i}) \end{cases} \quad (24)$$

$$\sigma^2_{\Delta F, BAR} = \frac{\langle f^2(x-C) \rangle_i - \langle f(x-C) \rangle^2_i}{N_i \langle f(x-C) \rangle^2_i} + \frac{\langle f^2(C-x) \rangle_j - \langle f(C-x) \rangle^2_j}{N_j \langle f(C-x) \rangle^2_j}$$

$$= \left( \frac{\langle f^2 \rangle_i}{N_i \langle f \rangle^2_i} - \frac{1}{N_i} \right) + \left( \frac{\langle f^2 \rangle_j}{N_j \langle f \rangle^2_j} - \frac{1}{N_j} \right) \quad (25)$$

$$= \sigma^2_i + \sigma^2_j$$

, where $f$ denotes the Fermi function and $\sigma^2_i$ is the variance contributed by the sample from state i. Using Fermi function to weight samples, we have the effective sample size

$$Q_{f,i} = \frac{N_i \langle f \rangle^2_i}{\langle f^2 \rangle_i} \quad (26)$$

. Still the dependence of the BAR variance on $Q$ is the same with the EXP one,

$$\sigma_i^2 = \frac{1}{Q_{f,i}} - \frac{1}{N_i} \tag{27}$$

. This is not unexpected, as the derivation of BAR follows the statistically optimal combination of perturbations in both directions. Hence, the statistical efficiency cannot be estimated with the effective sample size. The reliability of the results is mainly determined by the nature of weighting function. We note that the negative correlation between the effective sample size and the variance has been observed previously in a numerical study,[59] but the non-linear relationships in our results are found for the first time.

The formula of Gaussian approximated $Q_{f,i}$ for the Fermi weighting function can be easily obtained with the Fermi-Dirac-like integral. However, as in perturbation based theories $Q$ gives the same information with the variance of the free energy estimate, we do not find such formula useful.

Substituting Eq. (27) into Eq. (25), we combine the results in both directions as

$$\sigma_{\Delta F, BAR}^2 = \sigma_i^2 + \sigma_j^2 = \frac{1}{Q_{f,i}} - \frac{1}{N_i} + \frac{1}{Q_{f,j}} - \frac{1}{N_j} = \frac{1}{Q_{f,i}} + \frac{1}{Q_{f,j}} - \frac{2}{N} \tag{28}$$

, under the equal sample size rule $N_i = N_j = N$.

According to the range of the effective sample size from 1 to N given in Eq. (17), the variance in Eq. (27) is smaller than 1 and thus the variance of BAR estimate in Eq. (28) is smaller than 2, namely

$$\sigma_{\Delta F, BAR}^2 \in [0, 2) \tag{29}$$

. We can further get a general conclusion that without staging or stratification or model approximation (such as the Gaussian distributed energy difference), the variance of the free energy estimate from perturbation-based estimators always has an upper bound in alchemical transformation, and the upper bound is determined by the number of states we simulate or sample, and the number of perturbation we perform. Therefore, in FEP-based reweighting regimes, the best and the most reliable convergence-check method should be adding more windows, when the result does not show obvious change with further sampling in the existing windows. We note that TI, by contrast, does not have such a property (upper bounds). But integration methods also have their defects such as the bias introduced in integration.

From the definition of Eq. (25) we also know that

$$\sigma^2_{\Delta F,BAR} = \left( \int \frac{N_i N_j \rho_i \rho_j d\mathbf{q}}{N_i \rho_i + N_j \rho_j} \right)^{-1} - \frac{1}{N_i} - \frac{1}{N_j} = \frac{1}{N}\left( (O)^{-1} - 2 \right) \tag{30}$$

, where the probability density $\rho_i = \frac{e^{-U(\mathbf{q})_i}}{\int e^{-U(\mathbf{q})_i} d\mathbf{q}}$ and the overlap scalar $O = \int \frac{\rho_i \rho_j d\mathbf{q}}{\rho_i + \rho_j}$.[64] Then another interesting result can be obtained: The effective sample size is, actually, an estimation of phase space overlap, namely

$$O = \frac{Q_{f,i} Q_{f,j}}{N(Q_{f,i} + Q_{f,j})} \tag{31}$$

. From the range of the BAR variance in Eq. (29) and the relationship between the BAR variance and the overlap scalar in Eq. (30), we can also obtain the range of the overlap scalar, namely

$$\frac{1}{2N} \leq O \leq \frac{1}{2} \tag{32}$$

. This result can also be obtained from Eq. (31) and the range of the effective sample size or from the definition of the overlap scalar $O = \int \frac{\rho_i \rho_j d\mathbf{q}}{\rho_i + \rho_j}$ below Eq. (30). From the range of the overlap scalar, we know that if the overlap between two states is as small as 0.01, we need at least 50 independent samples to reach the correct answer. Too small sample size results in overestimated phase space overlap and fake convergence. In the large sample size regime the lower bound approaches zero. We also note that as TDV is linear dependent on the variance, TDV is also a nonlinear-function of the overlap scalar. Hence, as we already have variance-based criteria such as the overlap scalar and the TDV from which sufficient insights can be obtained, we do not find the effective sample size or other criteria useful.

The range of the overlap scalar in Eq. (32) shows sample-size dependent behavior. In the above discussion about the ranges of FEP variance and BAR variance we just prove the rough upper bounds, which by definition are sample-size independent. Here, follows the spirits of Eq. (32), we summary the range of these variance as the following equations.

$$\sigma^2_{\Delta F,FEP} = \frac{1}{Q_{EXP}} - \frac{1}{N} \in \left[ 0, \frac{N-1}{N} \right] \tag{33a}$$

$$\sigma^2_{\Delta F,BAR} = \frac{1}{Q_{f,i}} + \frac{1}{Q_{f,j}} - \frac{2}{N} \in \left[ 0, \frac{2N-2}{N} \right] \tag{33b}$$

. When N is a large number, the upper bounds approach the values we mentioned in the previous parts of the manuscript, namely $1(k_BT)^2$ for FEP and $2(k_BT)^2$ for BAR. The standard deviation of the free energy estimate is thus smaller than $k_BT$ for EXP and $\sqrt{2}k_BT$ for BAR.

We should note that all the above discussion focus on the statistical nature of the free energy estimator and thus can be directly extended to nonequilibrium free energy estimators of JI and CE. Further, considering all discussions above, in QM/MM corrections, if one still wants to use exponential averaging rather than the statistically optimal weighting, to narrow the distribution of energy difference (work), one can apply the staging strategy into equilibrium free energy simulation, or lengthen the duration of nonequilibrium transformation, or apply stratification in nonequilibrium simulation. A practical consideration is to narrow the standard deviation of the distribution to be smaller than 2. Still, one should remember that variance of the free energy estimate is underestimated and check the convergence carefully.

**Conclusion**

In this paper we present a theoretical explanation of the time-dependent behavior of various variance-based convergence criteria for perturbation-based estimators in free energy simulation. Our theoretical proof leads to the conclusion that for Gaussian distributed energy differences we should never use those effective sample size related criteria to determine the convergence of exponential average. Rather, the convergence of the distribution should be of much more importance. The reason is mainly that the Kish's effective sample size only provides a crude estimate of the sample size and in exponential averaging the phenomenon is exaggerated. The effective sample size from numerical timeseries data is overestimated compared with its asymptotically unbiased analytical value. The effective sample size is also proven to be a non-linear function of the variance of the EXP (and BAR) estimates. As normally the sample size is as large as 1000, the variance of the EXP estimate is approximately the reciprocal of $Q_{EXP}$. Thus $Q_{EXP}$ gives almost the same information as the variance. As $Q_{EXP}$ is significantly overestimated, the variance is orders of magnitude underestimated. The bounds of the effective sample size also lead to the bounds of the variance of the free energy estimate. The intrinsically underestimated variance is always smaller than $1(k_BT)^2$. Likewise, the variance of BAR estimate is smaller than $2(k_B T)^2$. The sample-size dependence of these variance estimates are given in Eq. (33). Hence, considering all the defects above we believe that in statistically inefficient EXP calculating the standard error to estimate the statistical uncertainty is meaningless, until the 'reliable' convergence is achieved. The effective sample size is also found to be a function of the overlap estimator in Eq. (31).

The number of samples needed for EXP convergence for Gaussian distributed energy difference with width $\sigma$ can be approximated from Eq.(20) in this paper. The minimum number of samples under large $\sigma$ can be interpolated with data under small $\sigma$ under the same convergence threshold. Further, when a desired accuracy is achieved, normally the variance of the free energy estimate and $Q_{EXP}$ are still biased, due to their dependence on higher-order terms. Thus, they play no role in bias detection and thus are useless in convergence assessment. Only with Gaussian approximation we can get some information from them.

As for obtaining unbiased free energy estimates, GEXP can only solve part of the question. For non-Gaussian situations we still have to use EXP. Poor convergence behavior is triggered by the inefficient weighting scheme. As the sample size required for convergence in EXP increases exponentially with the variance of the distribution, in a balanced scheme performing QM/MM simulation rather than only single point calculation and using bidirectional estimators should also be considered.

Finally, we should note that the current paper focuses on the two-state case, which is obvious due to the huge differences between computational cost of QM calculation and that of MM Hamiltonian. Multistate generalizations of the relationships found in this paper will be discussed along with representative examples in our following works.

**Appendix: Another way to derive Eq. (23) and Eq. (33)**

From Eq. (21) we know that to derive the upper bound of EXP in Eq. (23, 33), we just need to derive

$$\frac{\langle e^{-2x} \rangle}{N \langle e^{-x} \rangle^2} \leq 1 \tag{A1}$$

, or equivalently

$$\langle e^{-2x} \rangle \leq N \langle e^{-x} \rangle^2 \tag{A2}$$

. If we expand the ensemble averages in the above equation explicitly as

$$\langle e^{-2x} \rangle = \frac{1}{N} \sum_{i=1}^{N} e^{-2x_i} \tag{A3}$$

, and

$$\langle e^{-x} \rangle^2 = \frac{1}{N^2} \left( \sum_{i=1}^{N} e^{-x_i} \right) \left( \sum_{j=1}^{N} e^{-x_j} \right) \tag{A4}$$

$$N \langle e^{-x} \rangle^2 = \frac{1}{N} \sum_{i=1}^{N} e^{-2x_i} + \frac{2}{N} \sum_{j=2}^{N} \sum_{i=1}^{j-1} e^{-(x_i + x_j)} \tag{A5}$$

, the validity of Eq. (A1) and Eq. (A2) is rather clear, as the sum in Eq. (A3) is included in Eq. (A5) and $e^{-x}$ is always positive. Note that when N equals 1 the equality in Eq. (A1) and Eq. (A2) holds. The other part of Eq. (23, 33), the lower bound, requires

$$\langle e^{-2x} \rangle \geq \langle e^{-x} \rangle^2 \tag{A6}$$

, which is rather clear and does not need much derivation. Therefore, with Eq. (A2) and Eq. (A6), Eq. (23) and Eq. (33) can then be derived.

**Acknowledgement**

This work is supported China Scholarship Council. We thank Jiawei Zhang (University of Minnesota) for helpful discussion and Michele Parrinello (ETH Zuerich) for some discussion. We are grateful for the valuable and inspiring comments from the anonymous reviewers. Computer access to the CLAIX cluster of RWTH Aachen University and clusters of Forschungszentrum Juelich is gratefully acknowledged.

**Conflicts of interest**

There are no conflicts of interest to declare.

**Figure Captions**

**Figure 1.** Comparison between the numerical distribution generated and the analytical one (left) and an illustration of Gaussian distribution studied in this work with $\sigma$ from 1 to 10 (right).

**Figure 2.** Top: The effective number of samples as functions of $\sigma$ of the Gaussian distribution and the number of independent samples. We truncate the curve at $Q = 0.1$ as the sample size obtained from Eq. (16) should be larger than 1. Bottom: Its projections on $\sigma$ and its derivatives are also given. The numbers in the legend refer to the number of independent samples.

**Figure 3.** Comparisons of numerical ($Q_{EXP}$ from Eq. (16) ), theoretical and analytical ($Q_{GEXP}$ from Eq. (9) ) $Q - N$ curves. $\sigma =$ a) 0, b) 1, c) 2, d) 3, e) 4, f) 5. Here we can see that the theoretical line almost overlaps with the analytical line $Q_{GEXP}$ and the deviation between the numerical line $Q_{EXP}$ and the other two lines becomes larger with increased $\sigma$.

**Figure 4.** Convergence of EXP estimates (top) and GEXP estimates (bottom) in one exponential average trial under dimensionless $\sigma$ from 1 to 10. The larger $\sigma$ is, the harder a converged EXP estimate can be obtained.

**Figure 5.** The number of samples required to get certain accuracies and a confidence of 95%. The curves are colored by different required accuracies. Left: x axis as $\sigma$. Right: x axis as $\sigma^2$. The logarithm of the number of samples required to get a converged EXP estimate is found to be linearly dependent on the variance of the distribution in Eq. (19-20).

**Figure 6.** The slope of the $\ln N - \sigma^2$ curve as a function of the convergence threshold or required accuracy.

**Figure 7.** The timeseries of EXP variance in one trail for $\sigma$ from 0 to 10. We can see that the EXP variance is always smaller than 1 or $(k_B T)^2$, even in the small sample size regime. This indicates the variance of EXP is always underestimated and only when bias elimination is done the statistical uncertainty is meaningful.

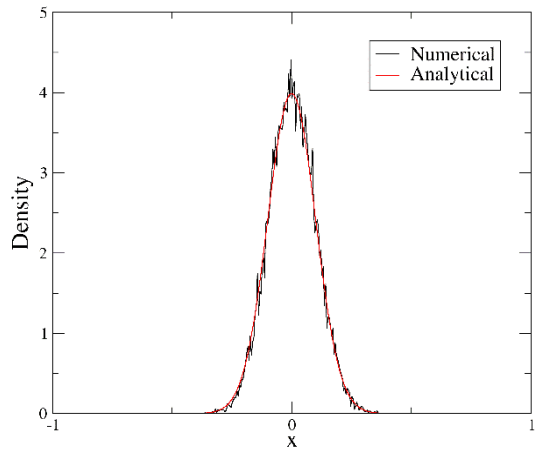 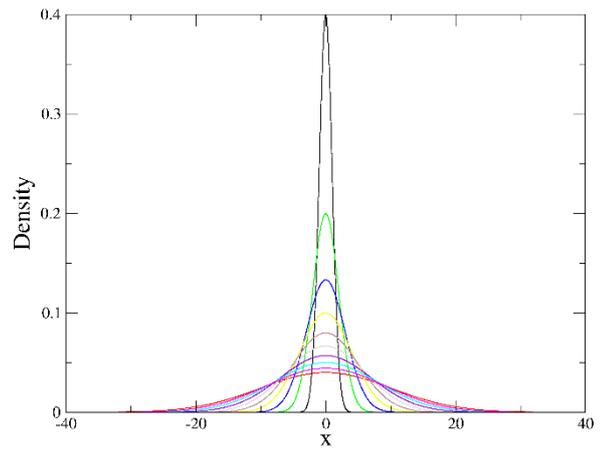

**Figure 1**

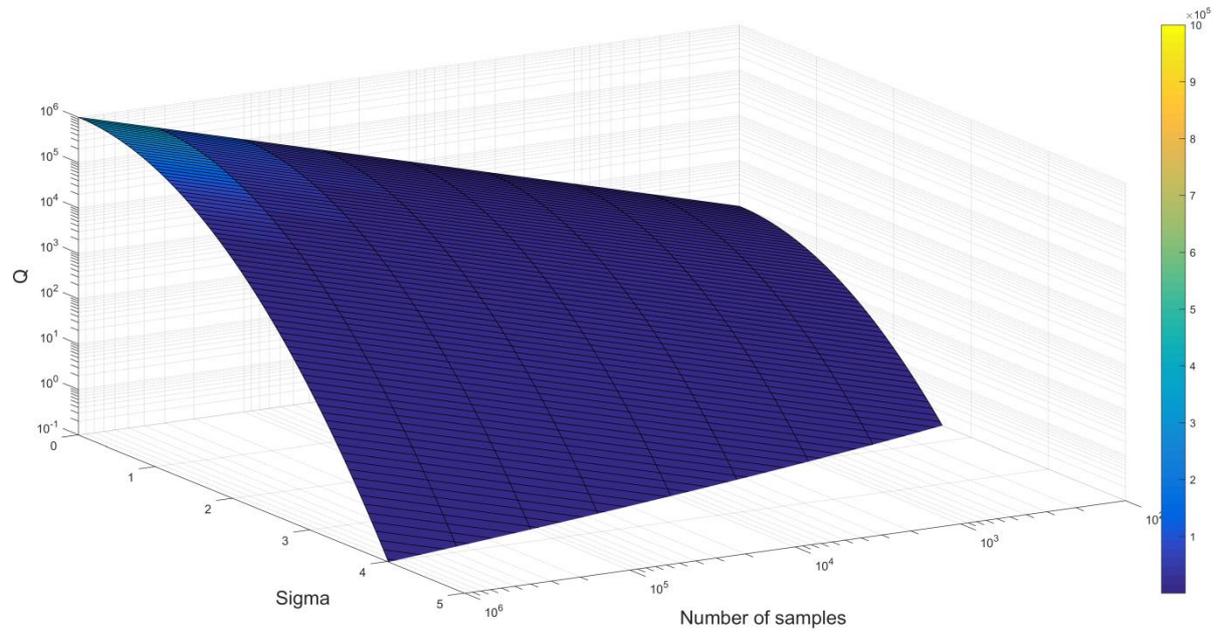
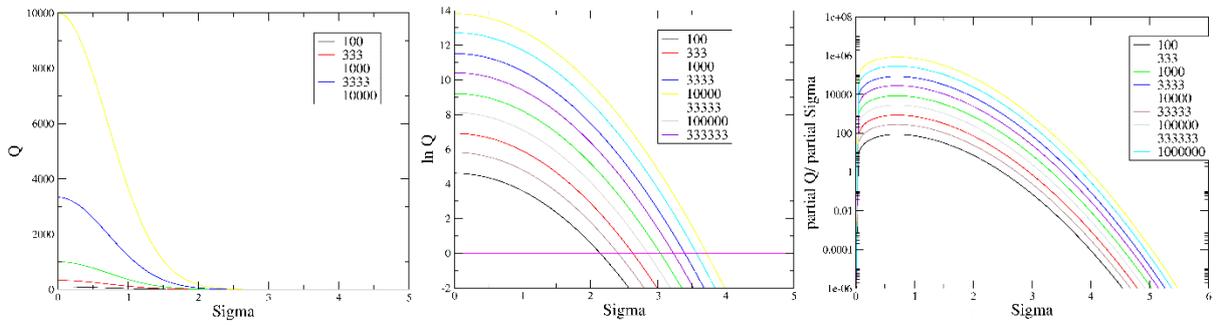

**Figure 2**

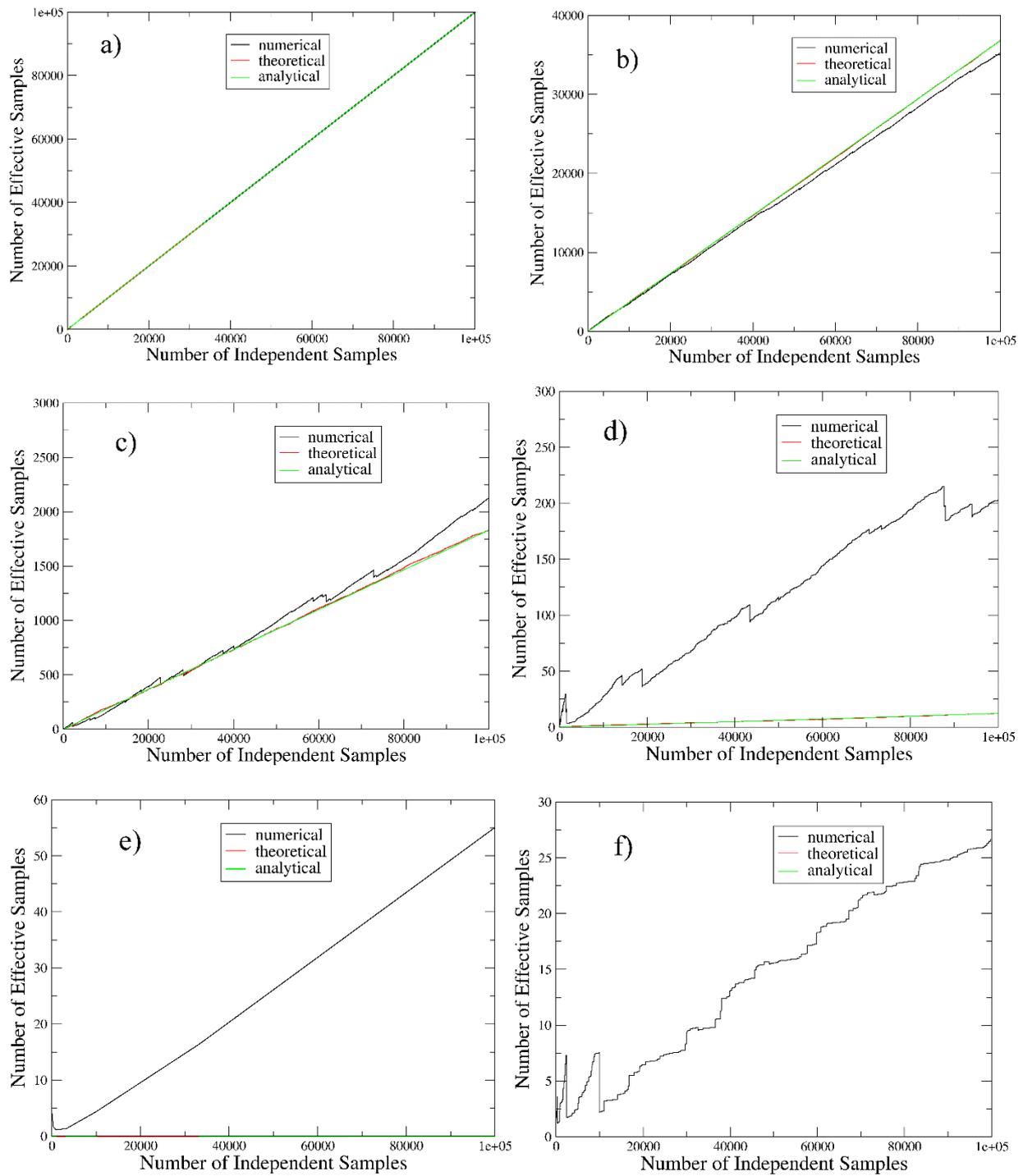

**Figure 3**

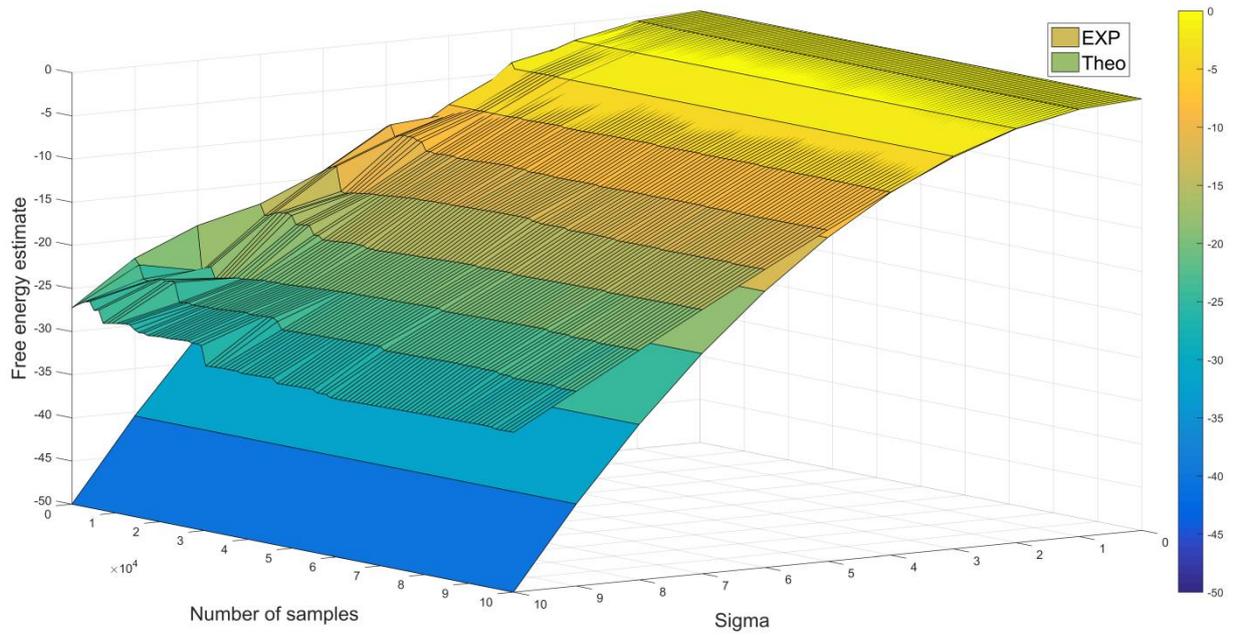

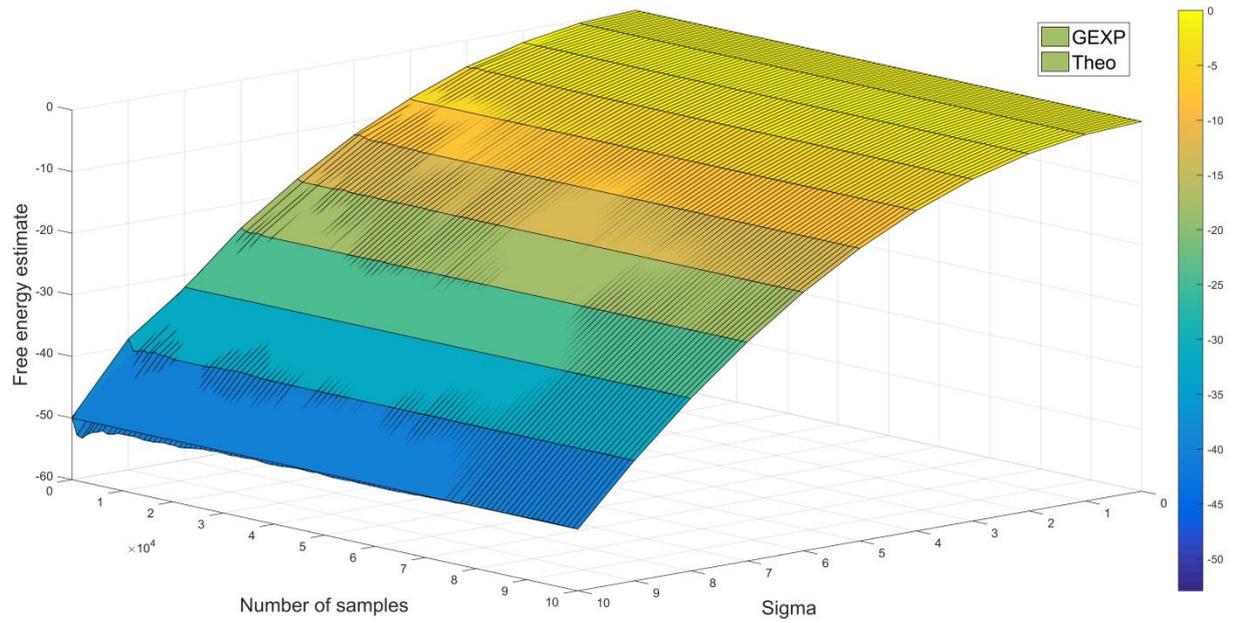

**Figure 4**

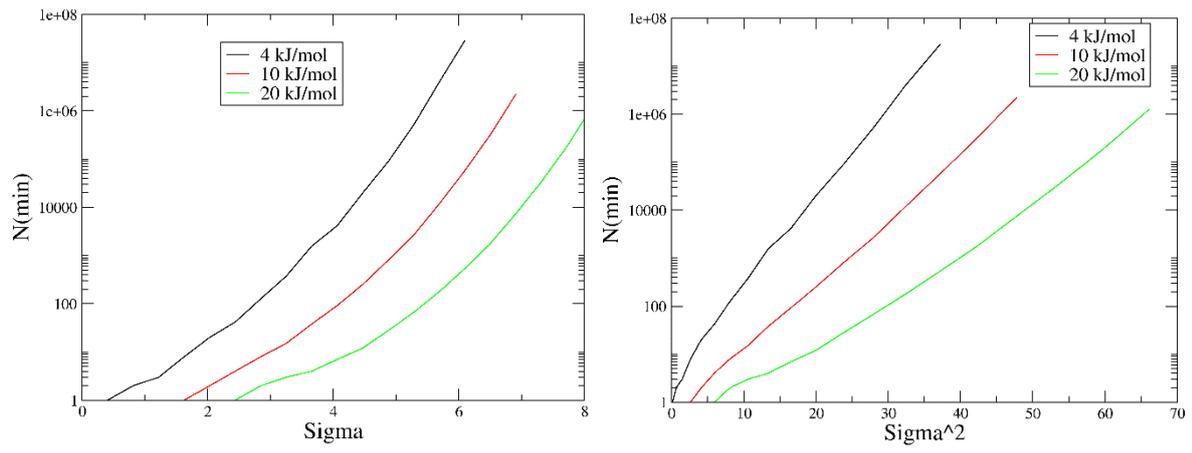

**Figure 5**

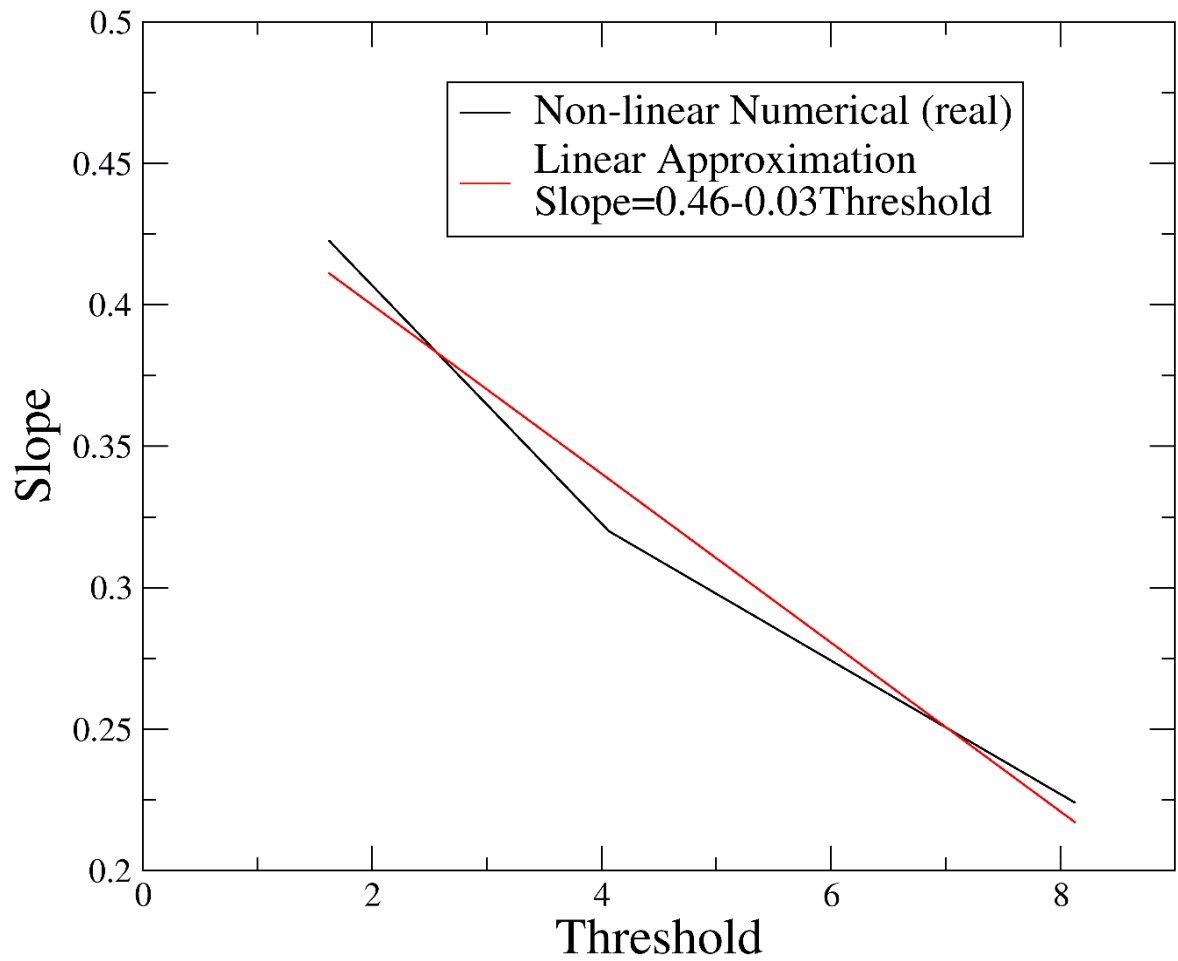

**Figure 6**

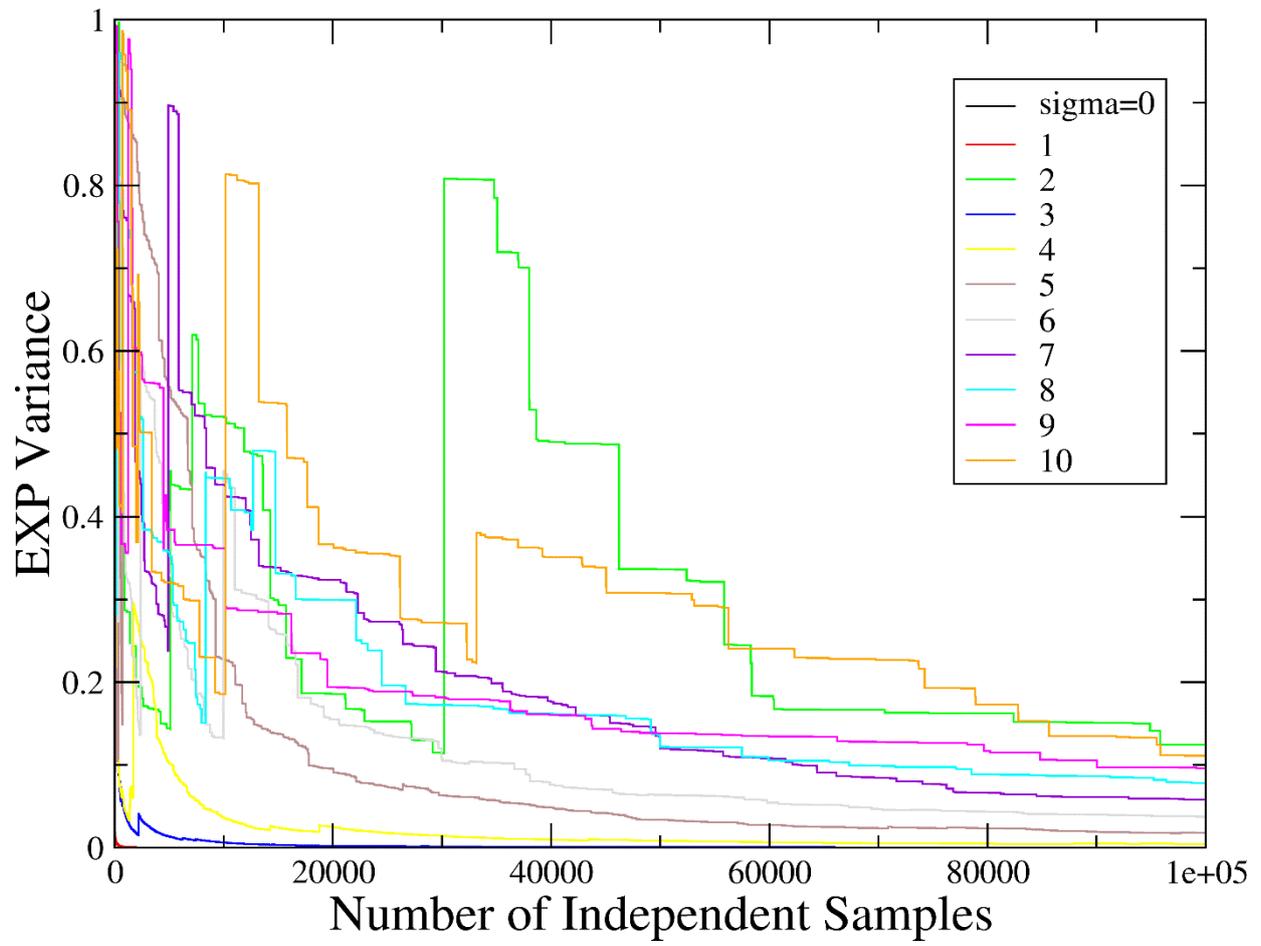

**Figure 7**